\documentclass[12pt,preprint]{aastex}
\usepackage{amsmath}
\usepackage{amssymb}
\usepackage{graphicx}

\usepackage{natbib}
\def\apj{ApJ} 
\def\mnras{MNRAS} 

\title{The Ionization of Accretion Flows in High Mass Star Formation: W51e2}

\author{Eric Keto\altaffilmark{1} and
Pamela Klaassen \altaffilmark{2}}
\affil{Harvard-Smithsonian Center for Astrophysics, 60 Garden
  Street, Cambridge, MA 02138, USA}
\affil{Dept. of Physics \& Astronomy, McMaster University, Hamilton, ON}
\altaffiltext{1}{keto@cfa.harvard.edu}
\altaffiltext{2}{klaassp@physics.mcmaster.ca}

\shorttitle{Ionized Accretion Flow in W51}
\shortauthors{Keto and Klaassen}

\begin{document}

\begin{abstract}
Previous observations show that the hypercompact HII region W51e2 is 
surrounded by a massive molecular accretion flow centered on the HII region.
New observations of the H53$\alpha$ radio recombination line made with
the VLA at  0.45 arc second angular resolution show a velocity gradient
in the ionized gas within the HII region of $> 500$ kms${-1}$ pc$^{-1}$
comparable to the velocity gradient seen in the molecular accretion flow. 
New CO line observations made with the SMA at arc second angular resolution
detect a molecular bipolar outflow immediately around the W51e2 HII region
and extending along the axis of rotation of the molecular flow.
These observations are consistent with an evolutionary phase for high 
mass star formation in which a newly formed massive star first begins to 
ionize its surroundings including its own accretion flow.
\end{abstract}


\section{Introduction}   

\subsection{HII regions within star-forming accretion flows}

Molecular line observations of several massive-star forming
regions show evidence for star-forming
accretion flows around small HII regions \citep{Guilloteau1983,
HoHaschick1986, Keto1987a, Keto1987b, Keto1988, HoYoung1996, ZhangHo1997, Young1998,
ZhangHoOhashi1998,  Sollins2004,  Sollins2005a,  Sollins2005b, Beltran2006}. 
These HII regions (W3(OH), G10.6-0.4, W51e2, G28.20-0.04,
and G24.78+0.08) are of hypercompact (HC) or ultracompact (UC) size and are bright
enough to require an ionizing flux at a level produced only by one or more O-type stars. 
Thus the presence and brightness of the HII regions at the centers of these flows 
indicate that the accretion is associated 
with the formation of  the most massive stars.  

The presence of the HII regions within molecular accretion flows
poses some simple yet interesting questions. How does the development
of an HII region affect accretion and the evolution of a young star?
Does the formation of an HII region prevent further accretion and  growth of the star?
How does
an HII region affect a pre-existing, bipolar outflow (BPO) of the type that
is typically
associated with the star-forming accretion flows of lower mass stars?

The interaction of HII regions with accretion flows and associated BPOs  occurs
in an evolutionary phase in massive star formation common to
all young O stars.  
Stellar structure calculations indicate that massive stars with accretion rates similar to those
observed around the HII regions listed above  ($10^{-3} - 10^{-2}$ M$_\odot$ yr$^{-1}$, 
\citep[][table 1]{Keto2002a})
should begin core nuclear burning at a mass well
below that of an O star 
\citep{StahlerShuTaam1980, StahlerShuTaam1981a, StahlerShuTaam1981b, PallaStahler1993,
BeechMitalas1994,NorbergMaeder2000,BehrendMaeder2001,KetoWood2006}. 
These calculations also indicate that the
temperature and luminosity of the accreting star
is about the same as a ZAMS star of equivalent mass. Thus 
an accreting star begins to produce an HII region once the star reaches a
mass, temperature and luminosity equivalent to early B, about 15 - 20 M$_\odot$. 
If the star is to accrete the additional mass to become an O star, then the
accretion must continue
past the mass when an HII region to forms.
Thus the interaction of accretion flows and HII regions is inevitable
in the early evolution of massive stars; however,
this interaction is currently not well understood.

Since bipolar outflows are invariably associated with
star-forming accretion flows of lower mass stars up to later-type B,
an  HII region that  develops within a pre-existing
molecular accretion flow must also
interact with the molecular bipolar outflow that we expect to be associated
with the accretion flow. 
There are few detections of outflows from young O stars. 
Young O stars are rare both because of the steep 
stellar mass spectrum and because of the short time that they
spend in the accretion phase. 
Furthermore, they are difficult to identify because they are deeply embedded in molecular
clouds and always at kpc distances. Maybe we have
not yet found enough examples, or maybe  
the HII regions that surround all young O stars modify or terminate the outflows.

\subsection{The ionized accretion flows}

Of the HII regions associated with molecular accretion flows, three have been
observed in radio recombination lines (RRL) with high enough angular resolution
to map the velocities of the ionized gas within the HII regions. 
All three HII regions show rotation in the ionized gas and one  
also shows infall. The similarities of the velocities
in the molecular and ionized phases 
suggest  that the flow is continuous across the ionization boundary. 
The UCHII region G10.6-0.04 which contains a 
small group of O stars is the best studied example, and the one that 
also shows an inward flow of the ionized gas
\citep{Keto2002a, KetoWood2006}. The G34.3+0.2 HII region 
shows rotation in its cometary shaped
ionized outflow \citep{Garay1986} that is perpendicular to a 
rotationally flattened molecular accretion flow
\citep{Keto1987b}.
More recently, \citet{Sewilo2007} report evidence
for rotation of the ionized gas within the G28.20-0.04 HII region in the 
same orientation and velocities as the rotation seen in the
``97 kms$^{-1}$" component of the NH$_3$(3,3) line previously mapped by
\citet{Sollins2005b}. In addition to these three, 
rotation of the ionized gas within the G45.07+0.13 HII region is
also reported in \citet{Garay1986}, but observations of the surrounding molecular
gas are inconclusive as to the presence of a molecular accretion flow 
\citep{Cesaroni1994, Wilner1996}. 

Inward flowing ionized gas has
been observed only in the large, cluster-scale HII
region G10.6-0.4. This HII region is the only one of the
three observed examples that has sufficient stellar mass 
inside the HII region (a few hundred M$_\odot$) that the
Bondi-Parker transonic
radius $R_g = GM/2c^2_s$  is large 
enough (several thousand AU) to be
observationally resolved at 
at the distances (6 - 8 kpc) of these three HII regions. 
The other two HII regions might also have inward flowing 
ionized gas, but if they contain only one O star or even a binary or triple,
then the inward flow will be on a length scale that is too small
to resolve observationally.

\subsection{The massive bipolar outflows}

There have been a few single dish telescope surveys that study 
BPOs in star forming regions with UC and HC HII regions.
\citet{ShepherdChurchwell1996}
and \citet{Klaassen2007} looked for BPOs
in massive star-forming regions with UC/HC HII regions.
\citet{Shirley2003} selected star-forming regions with water maser 
emission. Because water masers are associated with high mass stars, these 
regions also often
contain UC or HC HII regions.  These surveys
detect BPOs with mass outflow rates and energies that indicate that
they are driven by massive stars. However, these single dish observations
do not have the angular resolution necessary to determine
whether the BPOs are associated with the bright HII regions and therefore
O stars or with 
other nearby massive stars currently equivalent to ZAMS B without
HII regions.  
In order to determine whether a BPO is associated with 
a particular accreting massive star or an HII region,
we need higher resolution spectral line observations around the
HII region in molecules that
best trace bipolar outflows.

\section{New observations}

We chose to
observe the W51e2 HII region 
because of its well-studied molecular accretion flow \citep{Rudolph1990,
HoYoung1996, ZhangHo1997, ZhangHoOhashi1998, Sollins2004,
Young1998}.   In this paper we report on VLA observations of the H53$\alpha$ 
radio recombination
line (RRL) and on SMA observations of CO(J=2-1).  We observed the 
highest frequency RRL accessible by the VLA because the higher
frequency lines have lower optical depth, and we can observe the velocities
of the gas deeper inside the HII region. We observed the CO 
in a transition above the ground state 
and at $\sim 1^{\prime\prime}$ angular resolution to improve the likelihood of observing 
a BPO associated with hot molecular gas
near the HII region.

\subsection{Experimental setup}

Our H53$\alpha$ 
observations of the ionized gas in W51e2 were made at
the National Radio Astronomy Observatory
Very Large Array (VLA)\footnote{The National Radio Astronomy
Observatory is a facility of the National Science Foundation
operated under cooperative agreement by Associated Universities,
Inc.}.
The H53$\alpha$  line was observed in the 
C array at a resolution of 0.45$^{\prime\prime}$. 
The $^{12}$CO(J=2-1) line was observed at the Submillimeter Array (SMA)
\footnote{The Submillimeter Array is a joint project between the Smithsonian
Astrophysical Observatory and the Academia Sinica Institute of
Astronomy and Astrophysics, and is funded by the Smithsonian
Institution and the Academia Sinica.} in its extended
configuration, resulting in a spatial resolution  of $\sim 1''$. The CO line was
detected simultaneously with H30$\alpha$, and both sets 
of observations are described in \citet{KetoZhangKurtz2008}. 
The  
James Clerk Maxwell Telescope\footnote{The James Clerk Maxwell Telescope is 
operated by The Joint Astronomy
Centre on behalf of the Science and Technology Facilities Council of the United
Kingdom, the Netherlands Organisation for Scientific Research, and the National
Research Council of Canada.} was used to observe the larger scale structure of 
the molecular gas, and these observations were combined with 
the SMA observations using the non-linear
image combination program MOSMEM in the MIRIAD data reduction package 
(as described in Klaassen et al. in prep).

\subsection{The ionized flow}\label{ionized_accretion_flow}
Figure 1 shows the average velocity of the H53$\alpha$ line. 
The velocity gradient is
approximately 8 kms$^{-1}$ between 57 and 65 kms$^{-1}$ VLSR
with a position angle $-30^\circ$ (west of North). 
This velocity pattern is consistent with either rotation or outflow. 
We prefer the interpretation that
the H53$\alpha$ velocities indicate rotation for a couple of reasons.
First, the observed velocity gradient is perpendicular to the CO 
outflow. Bipolar outflows associated with lower mass stars 
are always oriented approximately
perpendicular to the rotation. Second, the interpretation of 
rotation is consistent
with the magnitude and direction of the rotational velocity gradient
observed in NH$_3$ and CH$_3$CN 
\citep{ZhangHo1997,  ZhangHoOhashi1998}.

The clearest signature of rotation in the molecular gas
is in figure 7 of 
\citet{ZhangHo1997} in the position-velocity diagram of
NH$_3$(3,3) along a line at position angle of 135$^\circ$,
and consistent (135 - 180 = $-55^\circ$)
with the position-angle of the velocity gradient
in the ionized gas ($-30^\circ$). 
We prefer to use the NH$_3$(3,3) line in this interpretation
rather than the (2,2) line
because the higher excitation line should derive from
gas closer to the hot HII region. 
However, there is considerable
uncertainty in locating the exact angle of the axis of rotation
using the molecular line observations. 
These observations map the flow at relatively large 
scales before the accretion
has spun-up by angular momentum
conservation to higher rotational velocities.
The observed infall velocities are comparable or greater
than the rotational velocities, and the rotational
component is difficult to extract. 
For example, comparison of
the position angle (-55$^\circ$) of the rotational 
gradient derived from the  NH$_3$ observations with
the position angle ($-70^\circ$) derived from CH$_3$CN 
\citep{ZhangHoOhashi1998}  suggests
an uncertainty of several tens of degrees.

Our observations
do not fully resolve the HII region, but the 1.3 cm continuum observation
of \citet{KetoZhangKurtz2008} 
indicates a
FWHM of $0.4^{\prime\prime}$ similar to the upper limit estimated by
\citet{ZhangHo1997} and \citet{Scott1978}. 
The radius,
$R$, associated with the velocity difference $\Delta V$, is 
uncertain because of  our low angular resolution. So we use 
the FWHM of the continuum emission as a characteristic size. 
Assuming that the observed velocities are rotational,
the mass of the star within the W51e2 region is estimated
as $M = \Delta V^2R/2G $.
Assuming a distance of 8 kpc, we derive a lower limit to 
the velocity gradient of $> 500$ kms$^{-1}$ pc$^{-1}$ and a 
dynamical mass of 
$V^2R/2G > 15$ M$_\odot$
This mass is roughly consistent
with a previous estimate of about 20 M$_\odot$ derived from the emission 
measure of the radio 
continuum \citep{Scott1978}. 

This estimated velocity gradient in the ionized gas
is similar to that measured in molecular gas.
\citet{ZhangHo1997} measure a velocity gradient in NH$_3$
of 500 kms$^{-1}$ pc$^{-1}$
centered around 58 kms$^{-1}$ VLSR. Owing to limited
angular resolution, the exact value of the
velocity gradient may be somewhat uncertain.
\citet{ZhangHoOhashi1998} derive a 
lower value (100 kms$^{-1}$ pc$^{-1}$) for the velocity gradient from
observations of CH$_3$CN.  
Nonetheless, the similarity in the magnitude and direction of the 
velocity gradients observed in the molecular and ionized  gas
suggests that there is a continuous flow of gas from the molecular to the 
ionized phase such that the ionized gas retains some memory of
the velocities of the molecular accretion flow. 

\subsection{The bipolar outflow}

Figure 2 shows the bipolar outflow in CO.
The orientation of the flow is along
the rotation axis of the ionized and molecular
accretion flows defined by the H53$\alpha$  RRL 
and NH$_3$ observations of \citep{ZhangHo1997}.
The outflow is seen as blue and red lobes in the maps of integrated emission.
In figure 2, the blue lobe is shown in the emission integrated from
20 to 44 kms$^{-1}$ (blue) and the red lobe from 68 to 92 kms$^{-1}$ (red). 
These ranges
are both 24 kms$^{-1}$ wide,
starting at velocities $\pm 12$ km s$^{-1}$ from the VLSR of the 
larger scale molecular gas (56 km s$^{-1}$). The blue
shifted lobe is much brighter than the red one and better collimated.
The better collimation of the blue flow results in a smaller volume. 
If the entrained mass is estimated from the volume and brightness, the
two flows have approximately the same mass (3 - 4 M$_\odot$) and
kinetic energy ($\sim 3\times10^{43}$ J), as calculated from zeroth 
and first moment maps of the outflow (Klaassen et al. in prep).

\section{Implications}

A conceptual model for the ionized flow in W51e2 may be constructed
as a composite of a couple of model calculations in the literature.
The expected structure (morphology) of an HII region that
develops within a large scale molecular accretion flow may be calculated from
ionization balance
\citep{Keto2007}. 
In particular, figure 1c of that paper
shows a nearly spherical HII 
region around a rotationally flattened molecular flow that may be relevant 
for W51e2. 
These calculations describe the structures of HII
regions but do not calculate the velocities or densities of the ionized gas.
Rather they assume an ionized flow based on the Parker
wind model \citep{Parker1958}. 
The velocities and densities
of the gas that flows into an HII from a rotationally flattened
molecular flow have been calculated in models of photo-evaporating
disks (PEDs) \citep{Hollenbach1994, Yorke1995, YorkeWelz1996, Lizano1996, 
Johnstone1998, Lugo2004}.  
However, these models do not calculate the structure of the molecular
accretion flow and HII
region for the types of molecular
accretion flows that are observed around HII regions. Rather they assume
the accretion flow is in the form of an infinitely thin disk. 
In the absence of a more complete calculation we may conceptually combine
the flow morphology from \citet{Keto2007} with the velocities and densities
from the PED models
to
interpret the the observations of W51e2.
The spin-up of the flow seen in the molecular line observations
suggests that the large scale molecular accretion
flow becomes progressively more flattened at smaller radii.  
If the flow is sufficiently dense at its mid-plane, it is able to resist ionization 
and extend into the HII region as described in \citet{Keto2007}. 
Across the top and bottom boundaries of 
the flattened molecular flow there is an ionization front that continuously
supplies gas to the HII region as described in the PED models.
Since there is no obvious force to 
immediately stop the rotation of the gas as it crosses the ionization
front, the ionized gas is also rotating. 
Observations of high frequency RRL such as the H53$\alpha$ are
dominated by the densest gas that is near the ionization front. 
The velocities of the H53$\alpha$ line
thus show the rotation in the ionized gas
coming off the rotating molecular accretion flow.

The density of the ionized gas decreases rapidly as it accelerates
off the ionization front around the flattened molecular accretion
flow. The exact rate of the density decrease 
depends on the geometry of  
the flow. The Parker wind model assumes
a purely spherical flow while the
PED models describe the flow off a flat, rotating disk. In both
cases, the density distribution is approximately exponential 
(hydrostatic) where the
flow velocities are subsonic and closer to a power law past the
transonic point. Observations in the radio continuum should thus
measure a rising spectral energy distribution (SED) consistent
with the steep density gradient in the
outflowing ionized wind. In the case of W51e2,
the observed SED may be reproduced by a model
with a density gradient scaling with radius as the $-2.5$ power
\citep{KetoZhangKurtz2008}.

The PED models show that the ionized gas flows along the rotation axis, 
and
the structures in \citet{Keto2007} show that this flow can be 
further channeled by the density structure of the molecular accretion
flow. 
This outflowing ionized gas could drive a molecular BPO, but this
BPO would be quite different in collimation and velocity structure 
from those BPOs
associated with the accretion flows of lower mass stars. At the moment,
the driver of the BPO that we see in CO in not known. It may
be the original BPO, but now flowing through the HII region, 
or it may be a remnant of a now terminated flow \citep{Klaassen2006}
or it might be driven by the ionized outflow.

\section{Conclusions}

Our new radio recombination line observations
of W51e2  
detect rotation in the ionized gas consistent with an
accretion flow, and our new CO line observations detect a
molecular bipolar outflow
along the rotation axis.

The observations are consistent with the structure expected
at the time that an accreting massive star begins to 
ionize its surroundings including the accretion flow responsible
for the formation of the star.

\bigskip\bigskip\noindent

\begin{figure}
\includegraphics[width=6.5truein, angle=90]{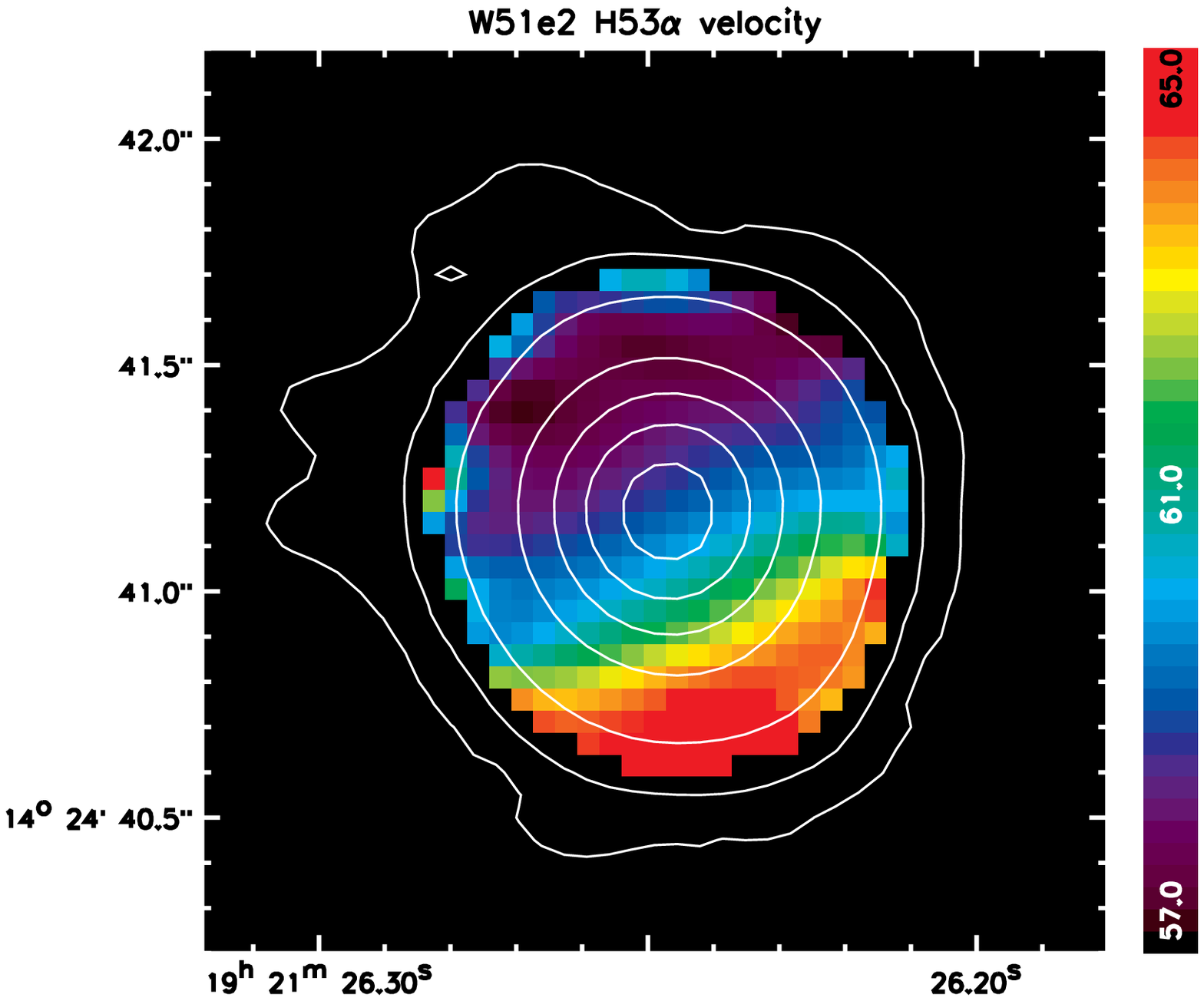}
\caption{Velocity of the H53$\alpha$ line in W51e2 in color. 
The color scale ranges from 57 to 65 kms$^{-1}$.  The contours show the 7 mm
continuum emission at 2, 4, 10, 30, 50, 70 and 90\% of the peak emission of
0.15 Jy/beam. Coordinates are in the B1950 epoch.}
\label{w51_h53}
\end{figure}

\begin{figure}
\includegraphics[width=6.5truein, angle=90]{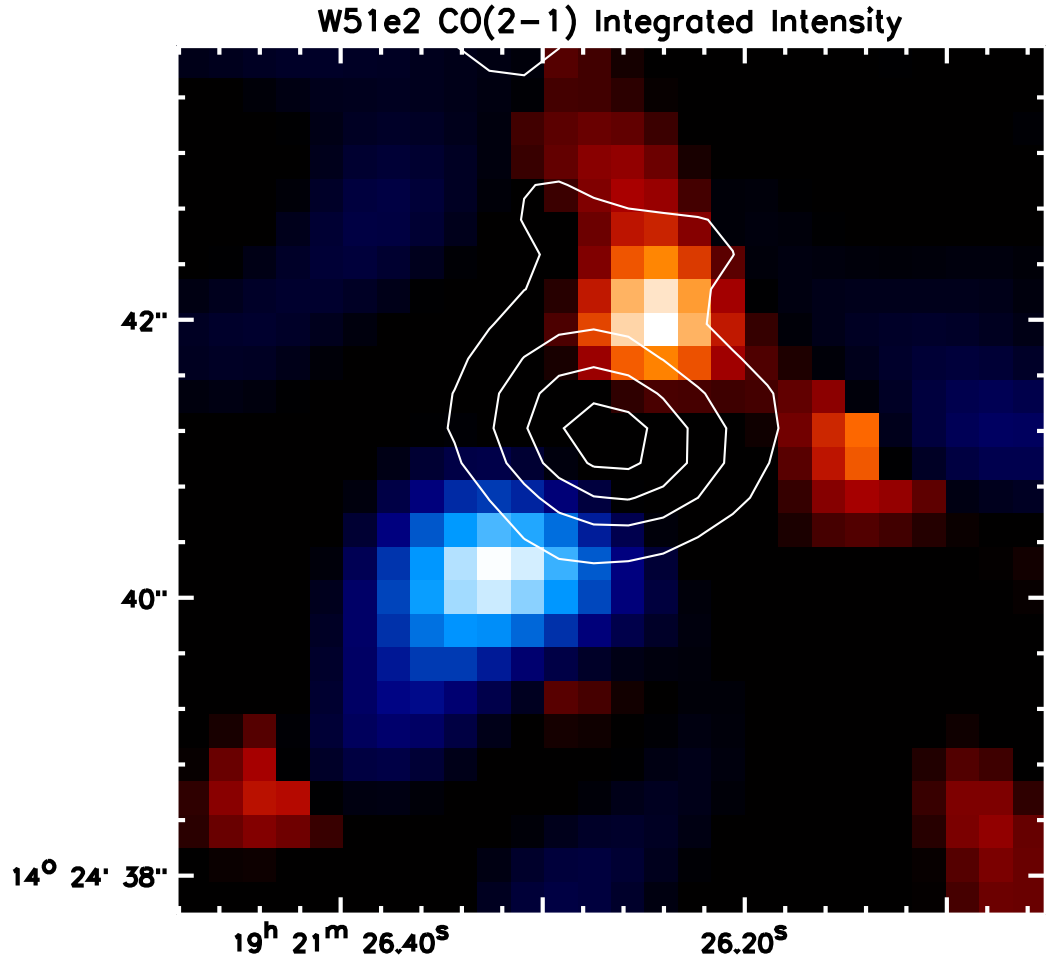}
\caption{ Integrated emission of the CO(2-1) line in color.
The emission from 20 to 44 kms$^{-1}$ is shown in blue and from 68 to 92 kms$^{-1}$
in red. The blue scale ranges from -17 to 47 K kms$^{-1}$.
The red scale ranges from -9 to 16 K kms$^{-1}$.
The contours show the 1 mm
continuum emission at 30, 50, 70 and 90\% of the peak emission of
2.1 Jy/beam. This map of W51e2 is on a larger scale than figure 1.
Coordinates are in the B1950 epoch.}
\label{co}
\end{figure}

\end{document}